# Towards combined optical coherence tomography and multi-spectral imaging with MHz a-scan rates for endoscopy

Madita Göb[1], Tom Pfeiffer[1] and Robert Huber[1]

[1]Institut für Biomedizinische Optik, Universität zu Lübeck, Lübeck, Peter-Monnik-Weg 4, 23562 Lübeck, Germany

## ABSTRACT

**Abstract:** We demonstrate a preliminary setup of a combined MHz-OCT and RGB narrowband reflection microscope and investigate the performance of the new RGB branch and different display modes of colored OCT data sets.

## 1. INTRODUCTION

With the development of multi-megahertz Fourier Domain Mode Locked (FDML) lasers, currently the fastest optical coherence tomography (OCT) systems are capable of live 3D imaging at video volume update rate [1]. Due to its high imaging speed and its ability to resolve biological microstructure in vivo, live 3D-OCT may enable improved clinical diagnostics, surgical guidance or guidance of biopsy. Standard OCT is very powerful in imaging tissue structure, however, its capabilities of detecting the molecular composition of a sample are limited. A multimodal approach of a combination of OCT with multi-spectral imaging allows to enhance the structural information of OCT by adding molecular information based on spectral reflection and thus potentially enables a variety of biomedical applications and diagnostic methods [2]. One application may be the identification of specific landmarks of a large sample based on spectral information. The tissue structure and depth related information may then be examined in particular using OCT. Furthermore, the spectral information can be used to render naturally looking OCT volumes, imitating the familiar impression of standard surgical microscope images.

In this paper, we demonstrate a setup using red, green and blue (RGB) visible laser diodes in combinations with a 1300 nm MHz-OCT engine. Comparable approaches were presented using standard speed OCT systems [3-4]. However, by using a multi-megahertz FDML laser, our system potentially allows colored live 4D-OCT at up to 25 volumes per second. We compare the performance of our RGB branch to a standard white light reflection microscope and investigate different display modes of the colored OCT data sets to present the huge amount of data to the user in the most clear and simple way.

## 2. METHODS

The schematic imaging setup is shown in figure 1. The RGB part of the system is realized by red, green and blue fiber coupled laser diodes, which are combined in an SMF28 fiber using a WDM (RGB25HA, Thorlabs). The combined light is collimated and then scanned by dual axis galvanometers on a scanning lens for focusing on the sample. The optical power on the sample was 1.5 mW at 637 nm, 0.85 mW at 520 nm and 0.93 mW at 465 nm. Light reflected back from the sample is extracted at the beam splitter and detected using an Avalanche Photodetector (APD110A/M, Thorlabs). The near-infrared light (NIR) of the OCT is coupled into the RGB path using a long pass dichroic mirror (DMLP950, Thorlabs), enabling simultaneous OCT and RGB imaging.

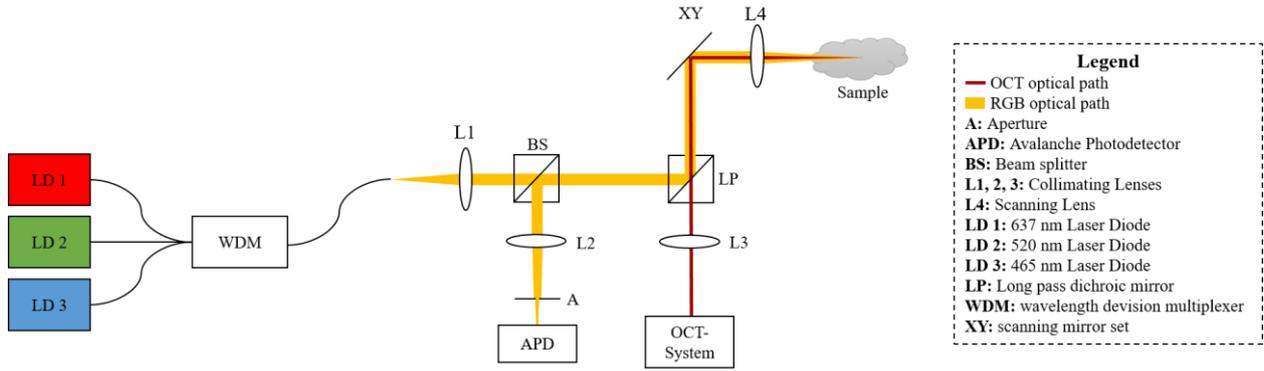

**Figure 1:** Basic detection setup for simultaneous OCT and spectral detection. All optical components are labled and described in the legend.

All data was acquired using a home-built 3.2 MHz FDML laser and a swept-source OCT system operating at a center wavelength of 1310 nm. For the OCT we used a 4 G samples/s analog to digital converter (ADC) yielding approximately ~1.5 billion voxels per second. However, the sampling rate of the RGB signal was limited to 1.25 M samples/s by our second ADC. In order to use one single photo-receiver for all three RGB channels the laser diodes were operated time-multiplexed, modulated line by line, to detect one color per line. A two-step interpolation of the RGB data followed in order to match the pixel-rate with the a-scan rate of the fast axis, as well as line-by-line. In the subsequent signal post-processing, the interpolated RGB images were rendered onto the surface of the OCT volumes.

## 3. RESULTS

Initial colored 3D-OCT volumes are shown in figure 2. The lateral resolution of the spectral system is ~25µm. Despite rather low signal levels partly due to high scanning speeds and non-optimized, uncoated optics, the colored surfaces of the OCT volumes in figure 2 appear very realistic. An image artifact located at the center of the RGB-images present in figure 2 (b) and 3 (b) is caused by reflections at the scanning lens. This problem only occurs when the light beam passes the scanning lens coaxially in the center.

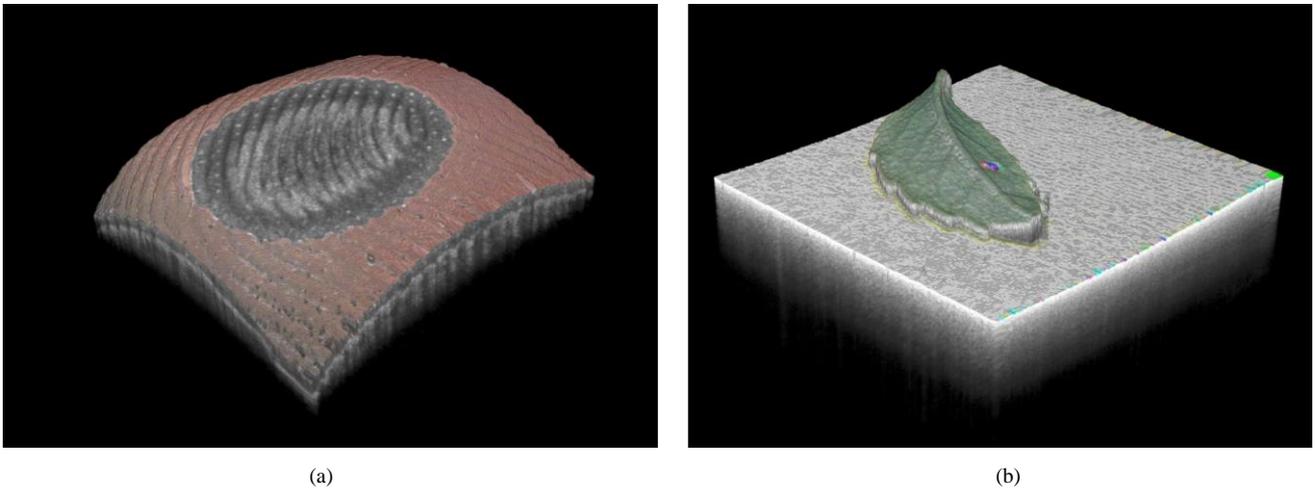

(a)      (b)

**Figure 2:** 3D OCT volumes of a fingertip (a) and green leave (b) with rendered RGB surface in different display modes using Fiji ImageJ.

The comparison of the captured RGB images with standard reflection white light microscope images shows that we can achieve very realistic color grading (see fig. 3 (a) and (b)). However, transverse resolution and contrast appear slightly inferior. In figure 3 (c-f) the spectral channels of an OCT-RGB en-face image displaying human skin with a nevus are shown separately. The contrast of the melanin-containing nevus is best using the green channel due to the absorption spectrum of melanin. In the NIR image (fig. 3 (c)) the nevus cannot even be recognized. This emphasizes the value of additional spectral information of this imaging technique.

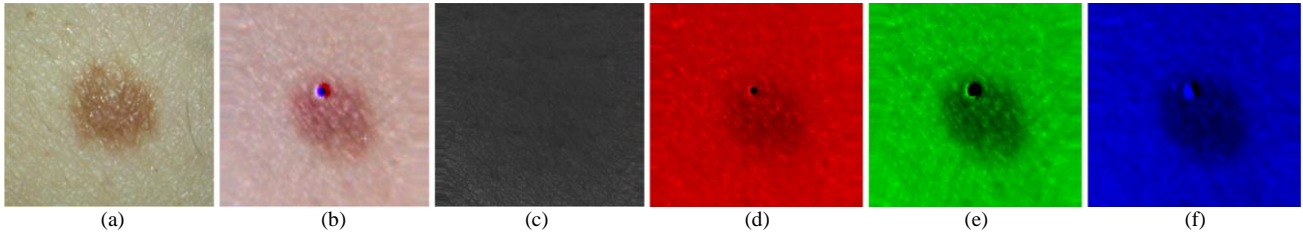

**Figure 3:** Image of a nevus displayed in different spectral channels. (a) Comparison image using standard reflection white light microscope, (b) RGB image, (c) OCT en-face image, (d-f) red, green, blue color channels. No artificial contrast correction methods were applied.

## 4. CONCLUSION AND OUTLOOK

We presented first results of a combined MHz-OCT and RGB narrowband reflection microscope. The additional information based on spectral reflection can not only be used to plastically visualize images in different display modes, but considering the value of multi-spectral reflectance characteristics, the system can be used to enhance the contrast of specific types of tissue. In the future, adding several more, selected wavelengths, the multimodal diagnostic value of the system may significantly increase. Future setups could for example include laser diodes at 660 nm and 940 nm enhancing the contrast of oxyhemoglobin and deoxyhemoglobin.

In contrast to previous works of combining OCT with spectral imaging, the benefit of our system is the exceptional imaging speed allowing live 3D imaging at video volume update rate. We are able to acquire OCT images with 3.2 MHz a-scan rate together with RGB. However, there are several problems, which are specific to MHz-OCT and the related high-speed detection. These are low signal levels due to high acquisition bandwidth, which also affects the electronic detection system. Further, due to synchronization issues, the spectral detection has not yet been integrated into the live 3D-OCT system. Our initial setup was also limited to line-wise interleaved detection modes by just one detector. We will discuss a new, improved setup and present its performance characterization.

Furthermore, the combined modalities show high promise for endoscopic applications since all modalities are fiber based. Our goal is the miniaturization of the preliminary setup enabling imaging of internal organs and tissue structures. An all fiber-based multispectral OCT setup would allow integration into a high-speed fiber scanning endoscope as presented by Schulz-Hildebrandt *et al.* [5].

## ACKNOWLEDGEMENTS

This project was funded by the European Union project ENCOMOLE-2i (Horizon 2020, ERC CoG no. 646669), the German Research Foundation (DFG project HU1006/6 and EXC 306/2; DFG EXC 2167), the German Federal Ministry of Education and Research (BMBF no. 13GW0227B: "Neuro-OCT"; BMBF no. 13N14665: „UltraLas") and by the European Union within Interreg Deutschland-Danmark from the European Regional Development Fund in the project CELLTOM.